\documentstyle[preprint,aps,floats,eqsecnum]{revtex}
\tighten

 \input amssym.def   
 \input amssym.tex

\input BoxedEPS
\SetRokickiEPSFSpecial  %% for dvips by Tom Rokicki, for
\HideDisplacementBoxes

\font\twelvemsb=msbm10 at 12pt
\font\ninemsb=msbm7 at 9pt
\font\sixmsb=msbm5 at 6pt
\newfam\Msbfam
\textfont\Msbfam=\twelvemsb
\scriptfont\Msbfam=\ninemsb
\scriptscriptfont\Msbfam=\sixmsb

\def\half{{\textstyle{1\over2}}}

\def\sixth{{\textstyle{1\over6}}} 
 
\newcommand{\fract}[2]{{\textstyle\frac{#1}{#2}}}

\def\beq{\begin{equation}}
\def\eeq{\end{equation}}
\def\bi{\begin{itemize}}
\def\ei{\end{itemize}}
\def\beqar{\begin{eqnarray}}
\def\eeqar{\end{eqnarray}}

\newcommand{\Aa}{\mbox{$\cal A\;$}}

\newcommand{\r}{\bbox{\rm r}}

\newcommand{\B}{\bbox{\rm B}}
\newcommand{\E}{\bbox{\rm E}}

\newcommand{\bv}{\bbox{\rm v}}
\newcommand{\bV}{\bbox{\rm V}}

\newcommand{\bj}{\bbox{\rm j}}
\newcommand{\ba}{\bbox{\rm a}}
\newcommand{\bk}{\bbox{\rm k}}
\newcommand{\bb}{\bbox{\rm b}}

\newcommand{\hn}{\hat{n}}
\newcommand{\ho}{\hat{\omega}}

\newcommand{\rmd}{{\rm d\null}}
\newcommand{\tr}{\mathop{\rm tr\null}}

\def\boldnab{\mbox{\boldmath$\nabla$}}

\def\boldsig{\mbox{\boldmath$\sigma$}}

\def\lra{\mathop{\hbox to .4in{\rightarrowfill}}}

\let\varkappa\kappa
\let\hat\widehat

\skip\footins = 14pt % was 12
\begin{document}

%% \nonfrenchspacing
%% \flushbottom

\title{Creation and evolution of magnetic helicity}

\author{R. Jackiw\footnotemark[1]}

\footnotetext[1] {\baselineskip=12pt This work is supported
in part by funds provided by  the U.S.~Department of Energy
(D.O.E.) under contracts
DE-FC02-94ER40818 and DE-FG02-91ER40676.\\
\null\hfill hep-th/9911072\qquad MIT-CTP-2919
\quad BUHEP-99-28 \qquad  November 1999\hfill\null}

\address{Center for Theoretical Physics\\ Massachusetts
Institute of Technology\\ Cambridge, MA ~02139--4307,
USA}

\author{So\kern 0.1em-\kern-0.18em Young Pi}

\address{Physics Department\\ Boston University\\
Boston, MA ~02215, USA}

\maketitle
\begin{abstract}\noindent
Projecting a non-Abelian $SU(2)$ vacuum gauge field -- a pure
gauge constructed from the group element $U$ -- onto a fixed
(electromagnetic) direction in isospace gives rise to a nontrivial
magnetic field, with nonvanishing magnetic helicity, which
coincides with the winding number of~$U$. Although the helicity
is not conserved under Maxwell (vacuum) evolution, it retains
one-half its initial value at infinite time.

\end{abstract}
\pagestyle{myheadings}
\thispagestyle{empty}

\markright{\small R. Jackiw and So\kern 0.05em-\kern-0.2em Young Pi --
\emph{Creation and evolution of magnetic helicity}}

\section{Introduction} \label{sec:Introduction}
It has been suggested that primordial magnetic fields can develop large
correlation lengths provided they carry nonvanishing ``magnetic helicity''
$\int
\rmd^3 r
\, 
\ba \cdot
\bb$, a quantity known to particle physicists as the Abelian,
Euclidean Chern-Simons term.  Here $\ba$ is an Abelian
gauge potential for the magnetic field $\bb = \boldnab
\times \ba$. If there exists a period of decaying turbulence in the early
universe, which can occur after a first-order phase transition, a magnetic
field with nonvanishing helicity could have relaxed to a large-scale
configuration, which enjoys force-free dynamics (source currents for the
magnetic fields proportional to the fields themselves) thereby avoiding
dissipation\cite{ref:1}.

In this paper we accomplish two things.  First we show that
configurations of $(\ba, \bb)$, with quantized helicity, arise
from vacuum
configurations of a non-Abelian $SU(2)$ vector potential.   The
quantization occurs because
$(\ba, \bb)$ wind and intertwine in an intricate manner.  We
relate this ``winding number'' of our Abelian fields to the
topological properties of the structures in a non-Abelian
$SU(2)$ gauge group.  Our construction is based on
$\Pi_3(S_3)$, which is relevant to a non-Abelian gauge
theory in 3-space.  [The construction can also be described
in terms of Hopf maps and $\Pi_3(S_2)$, which have already
appeared in the literature\cite{ref:2}; we show how our expressions
appear in that formalism.]  Only integer winding numbers
occur in the mathematical setting of $\Pi_3(S_3)$ or
$\Pi_3(S_2)$; yet we find that half-integer windings also
lead to interesting Abelian gauge fields.

Second, we study the time evolution of the magnetic
helicity.  Since $\frac{\rmd}{\rmd t} \int \rmd^3 r \, \ba
\cdot \bb = -2 \int \rmd^3 r \, \E \cdot \bb$ as a
consequence  of the definition for
the electric field, $\E$,  time variation is determined by the
specific physical situation that fixes $\E \cdot \bb$.  For
plasma physics or magneto\-hydrodynamics the projection of
$\E$ on $\bb$ is proportional to the resistivity of the
medium, and vanishes for infinite conductivity (zero
resistivity), see Sec.~\ref{sec:Conclusion}.  In that approximation, helicity
is conserved.  However, cosmological electromagnetic fields can
be expected also to experience evolution in vacuum, where
$\E \cdot \bb \ne 0$.  Therefore, it is interesting to
determine what happens to the magnetic helicity under
Maxwell evolution.  Specifically, we posit that at $t=0$ there
are magnetic fields with quantized helicity, and no electric
fields.  With this initial data, we solve the Maxwell equations
and find that at $t=\infty$ the helicity is precisely half its
initial value.  The calculation is carried out explicitly for two
interesting cases (integer and half-integer quanta) and then
a general argument is given, which requires a regularity
hypothesis.

In Section~\ref{sec:Helicity}, we describe how to construct  from
non-Abelian vacuum fields  Abelian gauge fields with quantized helicity.  In
Section~\ref{sec:Time}, we study time evolution. Concluding remarks
comprise Section~\ref{sec:Conclusion}, where we also speculate on the
applicability of these mathematical considerations to the $SU(2)\times
U(1)$ ``standard model", before and after its electroweak phase
transition. 

\section{Helicity of Gauge Fields}\label{sec:Helicity}

\subsection{Chern-Simons structures}

The Chern-Simons number of a non-Abelian gauge potential
$A_i^a$ is given by
\begin{eqnarray}
CS(A) &=& \frac{1}{16 \pi^2} \int \rmd^3 r \, \epsilon^{ijk} 
(A^a_i \partial_j A^a_k + \fract{1}{3} f^{abc} A^a_i A^b_j
A^c_k) \nonumber \\[1ex]
&=& -\frac{1}{8 \pi^2} \int \rmd^3 r \, \epsilon^{ijk}  \tr
(A_i \partial_j A_k + \fract{1}{3} A_i [A_j,A_k]) 
\nonumber \\[1ex]
&=& -\frac{1}{8 \pi^2} \int \tr
(AdA + \fract{2}{3} A^3)
\label{eq:2.1}
\end{eqnarray}
In the second equality, the $A_i$ are elements of the Lie
algebra with generators $T^a$
$$
A_i = A^a_i T^a  \ , \quad [T^a,T^b]=f_{abc}T^c  \ ,  \quad
\tr T^aT^b = -\delta_{ab}/2
$$
while form notation is used in the third equality: $A\equiv
A_i\, \rmd x^i$.  The normalization factor $\frac{1}{16 \pi^2}$
is chosen for later convenience, and is also maintained in the
Abelian limit, where only the bilinear part of (\ref{eq:2.1})
survives.  Thus our Chern-Simons quantity is $\frac{1}{16
\pi^2}$ times the magnetic helicity, which entered physics in
the work of Woltier\cite{ref:3} and was subsequently further
elaborated by many people, including Moffatt, Berger and
Field, as well as Arnold and Khesin\cite{ref:4}.  Particle physicists
have considered both  Abelian and non-Abelian
Chern-Simons terms on 3-dimensional Minkowski space in
studies of
$(2+1)$-dimensional gauge theories\cite{ref:5}.

To show how Abelian fields with nonvanishing and
quantized helicity arise from  non-Abelian vacuum
configurations, we recall first that under a gauge
transformation
\beq
A \to A^U \equiv U^{-1} AU + U^{-1} \rmd U
\label{eq:2.2}
\eeq
where $U$ is an element of the gauge group, the
Chern-Simons term transforms as 
\beq
CS(A^U)= CS(A) + \frac{1}{8\pi^2} \int \rmd (\tr \rmd
U\,U^{-1}A)  + \frac{1}{24\pi^2} \int \tr ( U^{-1}  \rmd U)^3
\label{eq:2.3}
\eeq
With sufficiently regular $A$ and $U$, the second term
on the right side integrates to zero\cite{ref:6}.  However, the last term,
also a total derivative -- although that is not apparent from
its formula (but see below) -- depends only on the
group element, and is not damped by any fall-off of $A$. 
Indeed its value is the winding number of $U$, which effects
a mapping from 3-space into the group.
\beq
W(U) = \frac{1}{24\pi^2} \int \rmd r ( U^{-1}  \rmd U)^3
\label{eq:2.4}
\eeq
$W(U)$ is an integer when $U$ is nonsingular for finite
$\r$ and tends to $\pm I$ at infinity, for then 3-space can
be compactified to the 3-sphere, and we assume that the
gauge group contains $SU(2)$.  Thus we are relying on
$\Pi_3 (SU(2))=\Pi_3(S_3)=$ integers.  It follows therefore
that the Chern-Simons number of a non-Abelian vacuum
gauge potential,   that is, one which is a pure gauge
\beq
\Aa= U^{-1}  \rmd U
\label{eq:2.5}
\eeq
is the winding number of $U$.

Let us now specialize to $SU(2)$.   Then $U$ involves the Pauli matrices
$T^a = \sigma^a/2i$.
\begin{eqnarray}
U &=& e^{\omega^aT^a} \nonumber \\[1ex]
&=& \cos f/2 - i \sigma^a \ho^a \sin f/2
\label{eq:2.6}
\end{eqnarray}
and the winding number is explicitly seen to involve a total
derivative\cite{ref:7}
\beq
W(U) = -\frac{1}{16\pi^2} \int \rmd \Big( \epsilon_{abc} 
\ho^a\, \rmd  \ho^b\, \rmd  \ho^c\,
(f-\sin f) \Big)
\label{eq:2.7}
\eeq
where $\ho^a$ is a unit vector in isospace:
$\omega^a = \ho^a f$.

\subsection{Constructing an Abelian gauge field}

Consider next the Abelian gauge potential 1-form $a$,
constructed from a non-Abelian vacuum configuration
$U^{-1}  \rmd U$ by projecting the latter on a fixed
 direction in isospace,
specified by a constant unit vector $\hn^a$
\begin{mathletters}
\beq
a=i \,  \tr \,  \hn^a \sigma^a U^{-1} \rmd U
\label{eq:2.8a}
\eeq
In components, this reads
\beq
a_i= \hn^a \Aa^a_i
\label{eq:2.8b}
\eeq
\label{eq:2.8}
\end{mathletters}%
where $\Aa^a_i$ is the vacuum (pure gauge) non-Abelian
potential
\beq
U^{-1} \partial_i U =\Aa^a_i \frac{\sigma^a}{2i}
\label{eq:2.9}
\eeq
which satisfies
\beq
\epsilon^{ijk} \partial_j \Aa^a_k = -\half \epsilon^{ijk}
\epsilon_{abc} \Aa^b_j  \Aa^c_k 
\label{eq:2.10}
\eeq
and carries the winding number
\beq
W(U) = -\frac{1}{96 \pi^2} \int \rmd^3 r \, \epsilon^{ijk}
\epsilon_{abc} \Aa^a_i  \Aa^b_j \Aa^c_k
\label{eq:2.11}
\eeq
Note that $\ba$ is {\bf not} an Abelian pure gauge.

We now show that the magnetic helicity of $\ba$ coincides
with $W(U)$. The magnetic helicity, with our normalization,
is given by
\begin{mathletters}\label{eq:2.12}
\begin{eqnarray}
H(a) &=& \frac{1}{16 \pi^2} \int \rmd^3 r \, \epsilon^{ijk} 
a_i \partial_j a_k \nonumber \\[1ex]
&=& \frac{1}{16 \pi^2} \hn^a \hn^b \int \rmd^3 r \, 
\epsilon^{ijk}  \Aa^a_i \partial_j \Aa^b_k 
\nonumber \\[1ex]
&=& -\frac{1}{32 \pi^2} \hn^a \hn^b \int \rmd^3 r \,
\Aa^a_i \epsilon^{ijk} \epsilon_{bcd} \Aa^c_j  \Aa^d_k
\label{eq:2.12a}
\end{eqnarray}
where (\ref{eq:2.8}) and (\ref{eq:2.10}) have been used.
But
$$
\epsilon^{ijk}  \Aa^a_i  \Aa^c_j \Aa^d_k = \sixth
\epsilon_{acd}
\epsilon^{ijk} \epsilon_{a'b'c'} \Aa^{a'}_i  \Aa^{b'}_j \Aa^{c'}_k
$$
Thus
\beq
H(a) = -\frac{1}{96 \pi^2} \int \rmd^3 r \, \epsilon^{ijk}
\epsilon_{abc} \Aa^a_i \Aa^b_j \Aa^c_k = W(U)
\label{eq:2.12b}
\eeq
\end{mathletters}
and we conclude that the magnetic helicity is quantized by
the winding number of the non-Abelian vacuum
configuration.

The form of the Abelian potential $\ba$ may be given
explicitly.  Parameterizing the unit vector  $\ho^a$ as
\beq
\ho^a = (\sin \Theta \cos \Phi, \sin \Theta \sin \Phi, \cos
\Theta)
\label{eq:2.13}
\eeq
and taking $\hn^a$ to point in the third
(electromagnetic) direction, we find
\begin{mathletters}
\beq
a=\cos \Theta\, \rmd f - (\sin f) \sin \Theta\, \rmd \Theta - 
(1-\cos f) \sin^2 \Theta\,  \rmd \Phi
\label{eq:2.14a}
\eeq
An alternate formula presents (\ref{eq:2.14a}) in ``Clebsch''
form, which will be useful later. (The Clebsch form for an
arbitrary 3-vector $\bV$ is $\bV=\boldnab \gamma+\alpha
\boldnab \beta$ where $\gamma, \alpha, \beta$ are three
scalar functions.)
\beq
a=\rmd (-2 \Phi) +2\Big(1-(\sin^2 f/2)  \sin^2 \Theta \Big)
\, \rmd (\Phi+\tan^{-1}[(\tan f/2) \cos \Theta]\Big)
\label{eq:2.14b}
\eeq
\label{eq:2.14}
\end{mathletters}%
The magnetic field is determined by the 2-form obtained from
(\ref{eq:2.14a})
\begin{mathletters}\label{eq:2.15}
\beq
\rmd a=\sin\Theta \Big[(1-\cos f)\,  \rmd f \, \rmd \Theta - 
2  (1-\cos f)\cos \Theta\,  \rmd \Theta \, \rmd \Phi + 
 (\sin f) \sin\Theta\,  \rmd \Phi \, \rmd f \Big]
\label{eq:2.15a}
\eeq
or from the Clebsch expression (\ref{eq:2.14b})
\beq
\rmd a=-2\, \rmd \Big( (\sin^2 f/2) \sin^2 \Theta \Big)
\,\rmd \Big( \Phi + \tan^{-1} [(\tan f/2) \cos \Theta]\Big)
\label{eq:2.15b}
\eeq
\end{mathletters}%
Finally the magnetic helicity becomes, according to
(\ref{eq:2.14a}) and (\ref{eq:2.15a})
\begin{mathletters}\label{eq:2.16}
\beq
H(a) = - \frac{1}{8\pi^2} \int (1-\cos f) \sin \Theta\, \rmd f \,
\rmd \Theta \, \rmd \Phi
\label{eq:2.16a}
\eeq
or from (\ref{eq:2.14b})  and (\ref{eq:2.15b})
\beq
H(a)=\frac{1}{4\pi^2} \int \rmd \Phi \, \rmd \Bigl( 
(\sin^2 f/2)  \sin^2 \Theta \Bigr)
\,\rmd \Big(\tan^{-1} [(\tan f/2) \cos \Theta]\Big)
\label{eq:2.16b}
\eeq
\end{mathletters}%
In the Clebsch parameterization, the magnetic helicity is
seen to involve integration of a total derivative and is
therefore given by a surface integral. 

The explicit Clebsch
expressions for $a$ and $\rmd a$ demonstrate that  one may find two
magnetic surfaces,
$S_n$ $(n=1,2)$, which satisfy
$\rmd a \, \rmd S_n = \bb \cdot \boldnab
S_n = 0$
\begin{eqnarray}
S_1 &=& (\sin f/2) \sin \Theta = c \nonumber \\[1ex]
S_2 &=& \Phi+ \tan^{-1}\Big[ (\tan f/2) \cos \Theta
\Big]=\phi_0 \nonumber \\[1ex]
&& c, \phi_0~{\rm constants}
\label{eq:2.17}
\end{eqnarray}
The intersection of these surfaces forms magnetic lines, that is,  integral
curves of $\bb$ that solve the dynamical system
\beq
\frac{\rmd \r (\tau)}{\rmd \tau} = \bb \Big(\r (\tau)\Big)
\label{eq:2.18}
\eeq
where $\tau$ is an evolution parameter for the dynamical
system.  Evidently, for our configuration this problem in
integrable, leading to curves given by
\begin{eqnarray}
\cos f/2 &=& \sqrt{1-c^2} \, \cos (\Phi-\phi_0) \nonumber
\\[1ex] 
\sin \Theta &=& \frac{c}{\sqrt{\sin^2(\Phi-\phi_0) + c^2
\cos^2 (\Phi-\phi_0)}}
\label{eq:2.19}
\end{eqnarray}

\subsection{Hopf Mapping}

It is known that the Abelian Chern-Simons term can be
related to the degree of the Hopf map, which is quantized
according to $\Pi_3(S_2)$\cite{ref:2}.  The construction of the relevant
vector potential proceeds in the following manner.  Consider
a complex spinor $u=\Big({u_1 \atop u_2}\Big)$ with $u^*u =
|u_1|^2 +|u_2|^2 = 1$.  Then $N^a \equiv u^*\sigma^a u$ is
a unit vector and the Hopf curvature is defined by
\beq
b^i = \frac{1}{4} \epsilon^{ijk} \epsilon_{abc} N^a \partial_j
N^b \partial_k N^c
\label{eq:2.20}
\eeq
Because $b^i$ is divergence-free, it can be written as the
curl of a potential $a_i$ given by $-i u^* \partial_i u$.  By
comparison with (\ref{eq:2.14b}) we find that 
our potential arises when $u$ is chosen as 
\beq
u = \left(
\begin{array}{c}
\displaystyle \sqrt{1-(\sin^2 f/2) \sin^2 \Theta}\,  
e^{2i\tan^{-1}[(\tan f/2) (\cos \Theta)]} \\[1ex]
\displaystyle (\sin f/2) \sin \Theta e^{-2i \Phi}
\end{array}
\right)
\label{eq:2.21}
\eeq
and the Hopf index coincides with the helicity of the
here-constructed $(\ba, \bb)$.

Evidently, the present expressions are awkward when
compared to those based on
$\Pi_3(S_3)$, presumably because the latter construction is directly
related to gauge fields.

\subsection{Explicit Expressions}

Henceforth we work with explicit expressions obtained by
identifying $\ho^a$ in (\ref{eq:2.6}) with the radial unit
vector $\hat{r}^a$ and taking $f$ to depend only on $r$.  This
further means that $\Theta$ in $\Phi$ in  (\ref{eq:2.13}) are
identified with polar and azimuthal angles of spherical
coordinates $\theta$ and $\phi$, with ranges $0\le \theta
\le \pi$ and $0\le \phi\le 2\pi$ respectively.  (A simple
generalization would be to identify $\Theta$ and $\Phi$
with integer multiples of $\theta$ and $\phi$.)  The
magnetic helicity  (\ref{eq:2.16}) then becomes
\beq
H=-\frac{1}{2\pi} \int^\infty_0  \rmd r \frac{\rmd}{\rmd r}
(f-\sin f) = -\frac{1}{2\pi}  (f-\sin f) \Big|_{r=\infty}
\label{eq:2.22}
\eeq
where we have taken $f(0)$ to vanish.  Observe that when
$\sin f(\infty)$ is nonvanishing $H$ is an
irrational/transcendental number.  When $f(\infty)$ is an
even integer multiple of $\pi$,  $\sin f(\infty)$ vanishes and $H$ is an
integer.  But note that also an  odd integer multiple of $\pi$ for $f(\infty)$
leads to vanishing $\sin f(\infty)$, and a half-integer value for $H$. 
Therefore it appears to us that configurations with
half-integer magnetic helicity are also singled out.  The half-integer
fields share further good properties with the integer-valued
configurations: we shall show below that only for these two the winding
number  does not change under the gauge transformation to the Coulomb
gauge. 

When reference is made back to the non-Abelian vacuum
configuration, which determines $\ba$, we see from
(\ref{eq:2.6}) that for the integer windings $U \lra\limits_{r
\to \infty} \pm I$, as expected.  For the half-integer ones, a
``hedgehog'' asymptote is attained: $U \lra\limits_{r \to
\infty} \pm i \boldsig \cdot \hat{r}$.

The vector potential $\ba$ and the magnetic field $\bb =
\boldnab \times \ba$ are neatly described by spherical
components, which are $\phi$-independent
\begin{mathletters}\label{eq:2.23}
\begin{eqnarray}
a_r &=& (\cos \theta) f'  \label{eq:2.23a} \\[1ex]
a_\theta &=& -(\sin \theta) \frac{1}{r} \sin f
\label{eq:2.23b} \\[1ex] 
a_\phi &=& -(\sin \theta)  \frac{1}{r} (1-\cos f) 
\label{eq:2.23c} 
\end{eqnarray}
\end{mathletters}%
\begin{mathletters}\label{eq:2.24}
\begin{eqnarray}
b_r &=&-2 (\cos \theta) \frac{1}{r^2}  (1-\cos f)
\label{eq:2.24a} \\[1ex] 
b_\theta &=&(\sin \theta)  \frac{f'}{r}  \sin f
\label{eq:2.24b} \\[1ex] 
b_\phi  &=&(\sin \theta) \frac{f'}{r}  (1-\cos f)
\label{eq:2.24c} 
\end{eqnarray}
\end{mathletters}%
(The dash denotes differentiation with respect to $r$.)
The Clebsch representation (\ref{eq:2.14b}), (\ref{eq:2.15b})
and (\ref{eq:2.16b}) gives a clear picture of the helicity. 
From (\ref{eq:2.16b}) we have
\begin{mathletters}\label{eq:new24}
\beq
H = -\frac{1}{8\pi^2} \int \rmd^3 r \, \partial_i \phi b^i = 
-\frac{1}{8\pi^2} \int \rmd^3 r \,  \frac{1}{r \sin \theta}
\Big( \frac{\partial}{\partial \phi} \phi \Big) b_\phi
\label{eq:new24a}
\eeq
Since $b_\phi$ is $\phi$-independent, the $\phi$
integration is trivial, leaving 
\beq
H = -\frac{1}{4\pi} \int^\pi_0 \rmd
\theta \int^\infty_0 r\, \rmd r \, b_\phi\Big|_{\phi=2\pi}
\label{eq:new24b}
\eeq
\end{mathletters}%
The integral is over the positive-$x$ $(x,z)$ half-plane and
$b_\phi$ is the toroidal magnetic field, perpendicular to that
plane.  So $H$ measures the flux of the toroidal magnetic
field through a half-plane.

Note that the functions occurring in our Clebsch parameterization are
multivalued, owing to the presence of the naked azimuthal angle $\phi$ and
the $\tan^{-1}$.  This is as it must be, because the helicity is
nonvanishing\cite{ref:new8}. Indeed  the multivaluedness of $\phi$ 
is responsible for the nonzero value of the $\phi$ ``surface'' integral
(\ref{eq:new24}), which reproduces the helicity.

When using $\ba$ in a description of
electromagnetic fields, we are effectively in
the $a^0=0$, Weyl gauge.  But also the vector potential must
be transverse since there are no sources.  However, the
potential in (\ref{eq:2.23}) is not transverse for general $f$. 
Transversality may be achieved in one of two ways.  We
may perform a gauge transformation, transforming $\ba$ to
a transverse expression $\ba^T$.  Alternatively, we may
choose the function $f$ so that $\ba$ is transverse.  We
discuss these two possibilities in turn.

The transversality condition affects only the poloidal
components of $\ba$ ($a_r$ and $a_\theta$), because the
toroidal component $(a_\phi)$ is $\phi$-independent.  By
using a gauge function proportional to $\cos \theta$
times a function of $r$, one can choose that function so that
the resulting, gauge equivalent potential is transverse.  The
components of the transverse potential then read
\begin{eqnarray}
a^T_r &=& (\cos \theta) F \nonumber \\[1ex]
a^T_\theta &=& -(\sin \theta) \frac{1}{2r} \Big(r^2 F\Big)'
\nonumber \\[1ex]
a^T_\phi &=&  -(\sin \theta)  \frac{1}{r} \Big(1-\cos
f\Big)  \nonumber  \\[1ex]
F(r) &\equiv& \fract{2}{3} \int^\infty_r \frac{\rmd
r'}{r^{\prime 2}} (f-\sin f) - \frac{4}{3r^3} \int^r_0 \rmd r' \,
r' (f-\sin f)
\label{eq:2.25}
\end{eqnarray}
Of course the magnetic field remains unchanged, but we
must still check the helicity integral: while the integrand is
gauge dependent, the integral changes only be a surface
term. It may be that the above gauge transformation 
contributes from the surface.  Indeed this happens for general $f$: one
finds
\beq
H(a^T) = H(a) + \frac{1}{6\pi} (1-\cos f) \sin
f\Big|_{r=\infty}
\label{eq:2.26}
\eeq
and  the helicity of the gauge-equivalent, transverse configuration differs
from the original expression.  However if, and only if, $f$ goes at
infinite $r$ to an even or odd multiple of $\pi$, the
gauge variance vanishes.  Thus both integer and half-integer
windings are stable against this gauge transformation, which renders the
potential transverse, but for other, irrational windings, the ``winding
number'' loses its meaning because it is gauge dependent\cite{ref:new9}.

\begin{figure}
$$
\BoxedEPSF{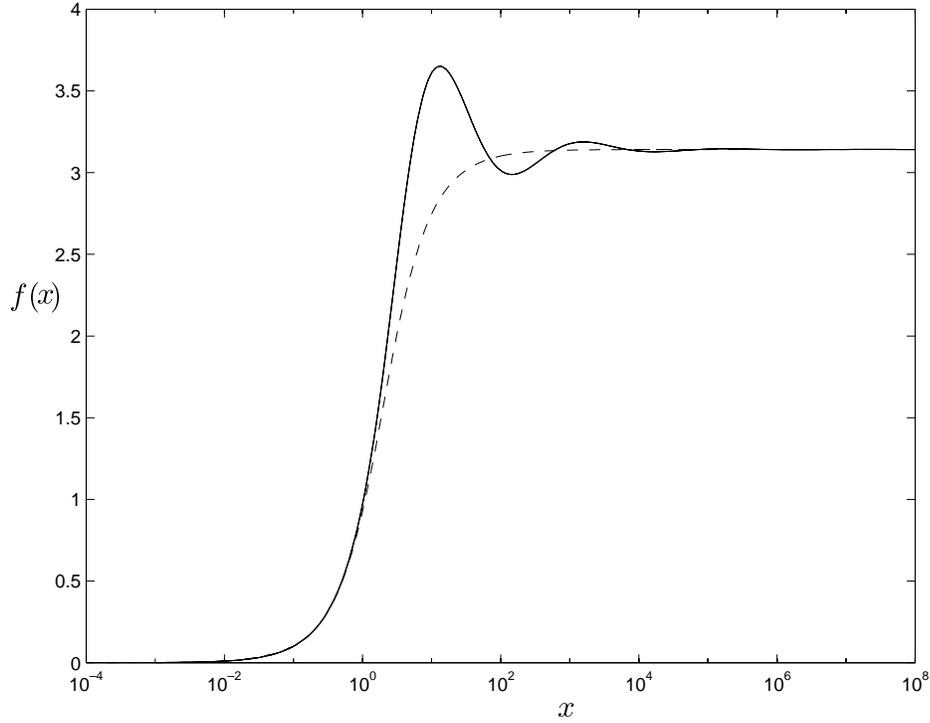 scaled 700}
$$
\caption{Profile of regular solution to (\protect\ref{eq:2.27}) (\emph{solid
line}); profile of $2\tan^{-1} x/2$, which possesses the same $x=0$ and
$x=\infty$ asymptotes (\emph{dashed line}).}
\label{fig:1}
\end{figure}

The other way to achieve transversality is to impose that
condition on the original configuration (\ref{eq:2.23}),
thereby determining $f$.  Transversality requires that $f$
satisfy
\beq
r^2 f'' + 2rf' - 2\sin f=0
\label{eq:2.27}
\eeq
Although this equation cannot be solved by elementary
functions, it is easily analyzed by analogy to a mechanical
problem where 	``time'' is $\ln r/r_0$.  One finds two
solutions that are regular at the origin, vanishing linearly with
$r$, and tending to $\pm\pi$ in an oscillatory manner for large~$r$. 
Moreover, the scale of $r$ is arbitrary: the solutions are a universal function
of $r/r_0$ and its negative; the positive solution is plotted in Fig.~\ref{fig:1}. 
Evidently this transverse potential necessarily corresponds to half-integer
winding\cite{ref:8}.

\section{Time Evolution}\label{sec:Time}

According to Maxwell's vacuum equations the transverse
vector potential satisfies the wave equation, which we
integrate subject to the initial condition that $\ba$ is given
at time $t=0$ by (\ref{eq:2.25}), with a definite choice for
$f$, and that $\frac{\rmd}{\rmd t}\ba$ is zero.  Thus at initial time, there
is no electric field and only the magnetic field (\ref{eq:2.24})
is present, maintained for $t<0$ by a steady current $\bj =
\boldnab \times \bb$.
\begin{eqnarray}
j_r &=& 2(\cos \theta) \frac{1}{r^2} (f-\sin f)' \nonumber
\\[1ex] 
j_\theta &=& -(\sin \theta)  \frac{1}{r} (f-\sin f)''
\nonumber \\[1ex]
j_\phi &=& (\sin \theta)  \Bigl( 
\frac{1}{r^2}[ r (1- \cos f)]' \Bigr)'
\label{eq:3.1}
\end{eqnarray}
and carrying energy~${\cal E} = \half \int \rmd^3r\, \bb^2$
\begin{eqnarray}
{\cal E} &=& \int_0^\infty \frac{\rmd r}{r^2} \sin^2 f/2 
\int_0^\pi \sin\theta\, \rmd\theta \Bigl[\Bigl(\sin^2 f/2 + r^2
f^{\prime 2}/4
\Bigr) + \cos2\theta \Bigl(\sin^2 f/2 - r^2
f^{\prime 2}/4\Bigr)\Bigr
]\nonumber\\ &=&
\frac{8\pi}3 \int_0^\pi \frac{\rmd r}{r^2} 
(1-\cos f)(1-\cos f + r^2 f^{\prime 2})
\label{3.2.new}
\end{eqnarray}

We determine the time asymptote for the helicity in two
cases 
\begin{mathletters}\label{eq:3.2}
\begin{eqnarray}
f&=& 2\tan^{-1} r/r_0 \label{eq:3.2a} \\[1ex] 
f &=& 4 \tan^{-1} r/r_0
\label{eq:3.2b}
\end{eqnarray}
\end{mathletters}%
The former, with $f \lra\limits_{r\to \infty} \pi$,
corresponds to half-integer winding; for the latter $f
\lra\limits_{r\to \infty} 2 \pi$,  one has integer
winding, also the energy density is spherically symmetric; see
(\ref{3.2.new})\cite{ref:8}.  The corresponding magnetic fields are
\begin{mathletters}\label{eq:3.3}
\begin{eqnarray}
\hbox{half-integer winding}  \qquad 
b_r &=& -4(\cos \theta) \frac{1}{r^2+r_0^2}
\nonumber \\[1ex] 
 b_\theta &=& 4(\sin \theta) \frac{r_0^2}{(r^2+r_0^2)^2}
\nonumber \\[1ex] 
 b_\phi &=& 4(\sin \theta) \frac{rr_0}{(r^2+r_0^2)^2} 
\label{eq:3.3a}\\[2ex]
\hbox{integer winding} \qquad 
b_r &=& -16(\cos \theta) \frac{r_0^2}{(r^2+r_0^2)^2}
\nonumber \\[1ex] 
b_\theta &=& 16(\sin \theta)
\frac{r_0^2(r_0^2-r^2)}{(r^2+r_0^2)^3}
\nonumber \\[1ex] 
b_\phi &=& 32(\sin \theta) \frac{rr_0^3}{(r^2+r_0^2)^3} 
\label{eq:3.3b}
\end{eqnarray}
\end{mathletters}%

The time-evolved fields are obtained by standard Fourier
transform techniques, and the helicity integral is evaluated
as function of time.  We find for the two cases
\begin{mathletters}
\begin{eqnarray}
H&=& -\half + \fract{1}{4} \frac{t^2}{r_0^2+t^2}
\Bigl( 1+\fract{2}{3} \frac{r_0^2}{r_0^2+t^2} \Bigr)
\label{eq:3.4a} \\[1ex]  
H &=& -1 + \half \frac{t^2}{r_0^2+t^2}
\Bigl( 1+ \frac{r_0^2}{r_0^2+t^2}  +
\frac{8r_0^6}{(r_0^2+t^2)^3} \Bigr)
\label{eq:3.4b}
\end{eqnarray}
\label{eq:3.4}
\end{mathletters}%
It is seen that for $t\to \infty$, $H$ attains one-half its value
at $t=0$.

While one would like to have an understanding how the
localization of the helicity changes with time, it does not
seem possible to pose such a question in a meaningful way,
because the helicity density, namely, the integrand that defines
$H$, is gauge dependent, and without invariant meaning. 
Indeed, as we have noted earlier, even the integrated
helicity can be gauge dependent when the fields and the
gauge function survive on the surfaces bounding the
integration region.  (This points to an
analogy with the energy in general relativity, whose density
is diffeomorphism-dependent.  Only the integrated quantity
is invariant and a unique value is determined only after
asymptotic conditions are prescribed.  Furthermore, the
energy may be presented as a surface integral and so also
can the helicity, when the Clebsch parameterization is used
for the gauge potential.)

The result that under Maxwell evolution $H$ decreases to
half its value at infinite time can also be understood from a
general argument.  In the Weyl-Coulomb gauge, which we
are using, the vector potential satisfies the wave equation,
which is uniquely solved with our posited initial conditions
by
\beq
\ba^T (t, \r) = \int \frac{\rmd^3 k}{(2 \pi)^3} \cos kt
e^{-i \bk \cdot \r} \tilde{\ba}^T (\bk)
\label{eq:3.5}
\eeq
Here $\tilde{\ba}^T (\bk)$ is Fourier transform of the
initial data (\ref{eq:2.25}) and $\frac{\rmd}{\rmd t}\ba^T (t, \r)=-\E (t, \r)$
vanishes at $t=0$.  It follows that 
\begin{eqnarray}
H &=& -i \int \frac{\rmd^3 k}{(2 \pi)^3} \cos^2 kt 
\epsilon^{ijk} k^i \tilde{a}^T_j(\bk) \tilde{a}^{T^*}_k(\bk)
\label{eq:3.6} \\[1ex]
 &=& \frac1{2i} \int \frac{\rmd^3 k}{(2 \pi)^3} (1+ \cos 2 kt)
\epsilon^{ijk} k^i \tilde{a}_j^T (\bk) \tilde{a}^{T^*}_k(\bk)
\label{eq:3.67}
\end{eqnarray}
By appealing to the Riemann-Lebesgue lemma, we can argue
that the term involving the cosine disappears at large $t$,
owing to destructive interference, leaving half the value at
$t=0$.  However, this step is justified provided the rest of
integrand is well behaved at $k=0$, which in turn depends
on the behavior of $\ba^T(\r)$ at large $r$.  A dimensional
estimate shows that a large-distance decrease of $\ba^T$
faster than $1/r$ is sufficient, which would mean that $\bb$
should decrease faster than $1/r^2$.  But (\ref{eq:3.3a})
exhibits a large $r$ behavior for $\bb$ of order $1/r^2$,
modulated by an angular factor.  Since our explicit
calculation supports the general argument, the angular
factor evidently provides sufficient large-$r$ damping.

\section{Conclusion}\label{sec:Conclusion}

Our investigation is based on the connection between the winding
number of non-Abelian gauge group elements and the Chern-Simons number
of Abelian gauge fields that are obtained from the former by
projection.  This mathematical fact suggests a physical scenario
for the $SU(2)\times U(1)$ ``standard model". Before its
(first-order) phase transition, its vacuum could be populated by
pure gauge, vacuum configurations $U^{-1} d U$ of the $SU(2)$
gauge group, which carry nonvanishing winding numbers.  After
the phase transition, one direction in isospace is identified with
electromagnetism, and the projection of the vacuum configuration
becomes a magnetic field with nonvanishing helicity. It remains
to be shown whether the above scenario is energetically stable and can be
justified on physical grounds. 

Another problem deserving further study concerns nontrivial
evolution of an initial configuration with magnetic helicity. Rather
than using the free Maxwell equations, one would rely on the
magnetohydrodynamical ones, which make use of Ohm's law to
express $\E$ in terms of $\bb$ and $\bj$. Its nonrelativistic
form is 
\beq
\bj = \eta (\E + \bv \times \bb)
\label{eq:4.1}
\eeq
where $\eta$ is conductivity and $\bv$ is the fluid  velocity,
taken to be divergenceless in a fluid of constant density. Inserting
this in the Maxwell equation
\beq
\frac{\partial \bb}{\partial t} + \boldnab\times \E = 0
\label{eq:4.2}
\eeq
produces an evolution equation for $\B$ 
\beq
\frac{\partial \bb}{\partial t} -  \boldnab\times (\bv\times\bb) 
 = -\frac1\eta \boldnab \times \bj \ . 
\label{eq:4.3}
\eeq
Further, approximating $\bj$ by $\boldnab\times \bb$, that is, ignoring
${\partial \E}/{\partial t} $ because it is negligible on the relevant
time scales,  converts the above into 
\beq
\frac{\partial \bb}{\partial t} -  \boldnab\times (\bv\times\bb) 
 = \frac1\eta \nabla^2 \bb 
\label{eq:4.4}
\eeq
which can be analyzed either with $\bv$ prescribed externally or
determined self-consistently by its Euler equation, with initial
$\bb$ of the form (\ref{eq:2.24}). An interesting choice for an external 
$\bv$ could be a transverse form that carries nonvanishing ``kinetic
helicity" $\int \rmd^3 r\, \bv \cdot (\boldnab\times\bv)$; for example, 
what one gets by taking $\bv$ in the form $\ba$ of (\ref{eq:2.23}) with $f$
solving  (\ref{eq:2.27}). For zero resistivity (infinite conductivity) the right
side of (\ref{eq:4.4})  is absent -- there is no dissipation. Then
$\E=-\bv\times\bb$ and magnetic helicity is conserved, since $\E\cdot\bb$
vanishes.
\bigskip

\centerline{\hbox to 1in{\hrulefill}}
\bigskip\noindent
Useful conversations with C.~Adam, A.~Polychronakos, and D.T.~Son are
acknowledged.

\end{document}